\documentclass[%
 a4paper,
 amsfonts,
 amsmath,amssymb,
 reprint,%
 onecolumn
]{revtex4-2}

\usepackage{graphicx}
\usepackage{dcolumn}
\usepackage{bm}
\usepackage{soul,color}
\usepackage[dvipsnames]{xcolor}
\usepackage{cleveref}
\usepackage[inline]{enumitem}
\usepackage[normalem]{ulem}

\begin{document}

\title{Sub-Kolmogorov Intermittency and Multifractal Dissipation in Multiphase Turbulence}

\author{Marco Crialesi-Esposito}
 \affiliation{DIEF, University of Modena and Reggio Emilia, via Vivarelli 10, Modena 41125, Italy}
\author{Aliénor Rivi\`ere}%

\affiliation{Sorbonne Universit\'e, CNRS, UMR 7190, Institut Jean Le Rond d'Alembert, F-75005 Paris, France}
\affiliation{Laboratory of Fluid Mechanics and Instabilities, Ecole Polytechnique F\'ed\'erale de Lausanne, CH-1015 Lausanne, Switzerland}

\author{Sergio Chibbaro}
\affiliation{LISN, AO team, B\^at 660 Universit\'e Paris-Saclay, Orsay Cedex 91405}
\affiliation{Inria Saclay - Tau team, B\^at 660 Universit\'e Paris-Saclay, Orsay Cedex 91405}

\date{\today}

\begin{abstract}
Multiphase turbulence displays stronger intermittency than its single-phase counterpart, yet the origin and geometrical organization of its most intense small-scale fluctuations remain poorly understood. Using direct numerical simulations of the incompressible Navier--Stokes equations with surface tension, we show that the local dissipative cutoff broadens strongly in the presence of interfaces, with dissipative events extending deep into the sub-Kolmogorov range. These events are spatially concentrated around topology-changing interfacial regions, namely breakup and coalescence. A multifractal analysis of the dissipation field further reveals that, while the spectrum above the Kolmogorov length, $\eta_K$, remains close to the single-phase case except for the most singular tail, the near- and sub-Kolmogorov range develops a markedly broader singularity spectrum supported on sparse intense structures. 
Our results show that breakup and coalescence do not simply perturb turbulence locally, but imprint a distinct multifractal organization on dissipation in multiphase turbulence.
\end{abstract}

\maketitle

A key challenge in multiphase turbulent flows is to understand the dynamics at the smallest active scales, which ultimately control droplet-size distributions in many geophysical and industrial processes~\cite{garrett2000connection,deane2002scale,lasheras2002atomization}. 
In contrast to single-phase turbulence, multiphase flows involve deformable and topology-changing interfaces, which profoundly alter the transfer of energy across scales. In particular, recent studies have shown that surface tension mediates a scale-dependent exchange between turbulent kinetic energy and interfacial energy, thereby modifying the classical picture of the cascade~\cite{Perlekar2019,Crialesi-Esposito2022,crialesi2023interaction,Saeedipour2023,Thiesset2025}. 

A direct consequence of this coupling is an enhancement of small-scale intermittency. In single-phase turbulence, intermittency is reflected in the strong non-Gaussian fluctuations of velocity increments and dissipation, associated with a breakdown of scale invariance in the cascade process~\cite{kolmogorov1962refinement,sreenivasan1997phenomenology,Benzi2023}. 
In multiphase turbulence, these effects are significantly stronger: turbulent emulsions display enhanced anomalous fluctuations, and recent evidences indicate that intense dissipation events are closely tied to the formation of small droplets and to interface-mediated velocity gradients~\cite{Crialesi-Esposito2023,Crialesi-Esposito2024,riviere2024,Fan2025}. 

Despite these advances, the physical origin and geometrical organization of small-scale fluctuations in multiphase turbulence remain poorly understood. In particular, a key open question is that, while breakup and coalescence are known to generate strong local gradients and to reshape the small-scale flow~\cite{ni2024deformation}, it is still unclear how these topology-changing events are reflected in the statistics of the dissipative field and in the fluctuations of the local dissipative cutoff~\cite{Crialesi-Esposito2024,eggers1997nonlinear}.

Addressing this question requires probing the smallest dynamically active scales, where a significant fraction of mixing and dissipation takes place~\cite{sreenivasan2019turbulent}. This issue naturally calls for a multifractal description. 
Intermittency is intimately connected with the spatial organization of dissipation, which is known to concentrate on highly intermittent, non-space-filling sets. This observation motivated the development of the multifractal formalism~\cite{benzi1984multifractal,Paladin1987}, later established as a predictive framework for turbulence~\cite{frisch1996turbulence,Boffetta2008}. In this framework, velocity increments and coarse-grained dissipation are characterized by fluctuating local scaling exponents, associated with a singularity spectrum that encodes the geometry of the most intense events~\cite{Meneveau1987,Meneveaut1991}. While this approach has proved highly successful in single-phase turbulence, its extension to multiphase turbulent flows is, to our knowledge, still missing.

In this paper, we put forward the statistical description of the dissipation field and its link with the interface by combining an analysis of fluctuating dissipative scales with the singularity spectrum of the dissipation field.
We show that the widening of the dissipative range is directly linked to enhanced intermittency and that the most singular fluctuations are statistically tied to topology-changing interfacial events, namely breakup and coalescence. This provides a direct connection between interfacial dynamics, local Reynolds-number fluctuations, and the extreme small-scale activity of multiphase turbulence.

Our analysis is based on direct numerical simulations of the incompressible Navier--Stokes equations with surface tension. The interface dynamics is captured with a Volume-of-Fluid method~\cite{tryggvason2011direct} using the code FLUTAS~\cite{Crialesi-Esposito2023-Flutas}. 
The numerical experiments are designed to provide 
a 
high resolution of the dissipative range.
All simulations are performed in a triply periodic box of size $L=2\pi$, with turbulence maintained by large-scale forcing at the box size (see Supplemental Information). 
The computational domain is discretized on a grid of $512^3$ points. 
We consider two configurations: a single-phase reference flow and a multiphase flow. In the single-phase case, we obtain a Taylor Reynolds number of $Re_\lambda \approx 50$ and a Kolmogorov length scale of $\eta_K=(\nu^3/\bar{\varepsilon})^{1/4} \approx 6.8\,\Delta x$, where $\Delta x$ is the grid spacing, $\nu$ the kinematic viscosity,  and $\varepsilon = 2\nu S_{ij}S_{ij}$ is the local energy dissipation rate (the overbar denotes time-space average), with $S_{ij}=(\partial_i u_j+\partial_j u_i)/2$.
In the multiphase case, the carrier and dispersed phases have identical density and viscosity. We consider a dispersed-phase volume fraction $\psi=0.1$, corresponding to a large-scale Weber number
$\mathrm{We}(d_v)={\rho\,\bar{\varepsilon}^{2/3} d_v^{5/3}}/{\sigma}=690$,
where $d_v = 2\left( \frac{3\psi L^3}{4\pi} \right)^{1/3}$ is the volumetric diameter and $\sigma$ is the surface-tension coefficient. This value leads to a Kolmogorov-Hinze scale, $d_h$, at which surface tension forces balance, in average, inertial forces, defined by $\mathrm{We}(d_h) \approx 1$, of $d_h = 4\eta_K$. Matching density and viscosity between the two phases allows us to isolate the effect of interfacial dynamics and surface tension on the small-scale statistics. 

\begin{figure}[h]
	\centering
    \includegraphics{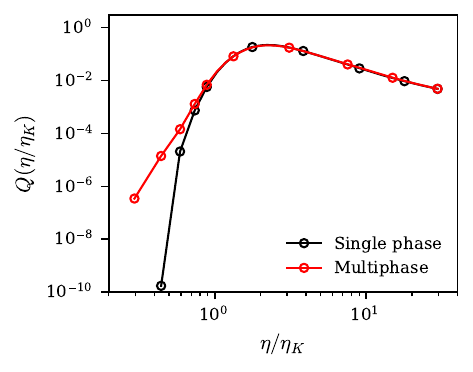}
	\caption{
    Probability density function $Q(\eta/\eta_K)$ of the local dissipative scale $\eta/\eta_K$ defined by the condition $Re(\eta)\sim 1$ in \Cref{eq:localRe}~\cite{Schumacher2007a}, obtained in single phase (black line) and multiphase (red line) simulations. 
    }
	\label{fig:qeta}
\end{figure}
A central consequence of intermittency in single phase turbulence is that the dissipative cutoff is not fixed at the Kolmogorov scale $\eta_K$, but fluctuates strongly in space and time, with rare events extending deep into the sub-Kolmogorov range~\cite{Paladin1987,frisch1996turbulence,Schumacher2007a,Biferale2008}.
Within the multifractal framework, \citet{Paladin1987} related velocity fluctuations to a scale-dependent local Reynolds number of order unity,
\begin{equation}
	\mathrm{Re}(\textbf{x}, \eta) \equiv \frac{\eta\,|\delta u_\eta(\textbf{x})|}{\nu} \sim 1,
	\label{eq:localRe}
\end{equation}
where $\delta u_\eta(\textbf{x}) = u_i(x_i+\eta,t)-u_i(x_i,t)$ denotes a local one-dimensional longitudinal velocity increment across the scale $\eta$. This condition follows from a balance between convective and viscous time scales, or equivalently from a local energy budget~\cite{Paladin1987,Yakhot2005}, and defines an instantaneous local dissipative scale.
The probability density $Q(\eta)$ of the scales satisfying $\mathrm{Re}(\textbf{x}, \eta)\sim 1$ therefore characterizes the distribution of the local dissipative cutoff-scale and quantifies how far intermittent events penetrate into the dissipative range~\cite{Schumacher2007a}. \Cref{fig:qeta} shows $Q(\eta)$ for the present simulations. Compared with the single-phase case, the multiphase flow displays a markedly broader distribution. In particular, the left tail is substantially enhanced, indicating a much higher probability of events with $\eta \ll \eta_K$, \textit{i.e.} intense fluctuations are found deep into the sub-Kolmogorov range. 
Altogether, these results provide direct evidence that interfaces amplify the intermittency of the dissipative dynamics.
This immediately raises two related questions: what generates these intense sub-Kolmogorov events, and how do they affect the global flow intermittency? 

\begin{figure*}[t]
	\centering
	\includegraphics[width=0.98\linewidth]{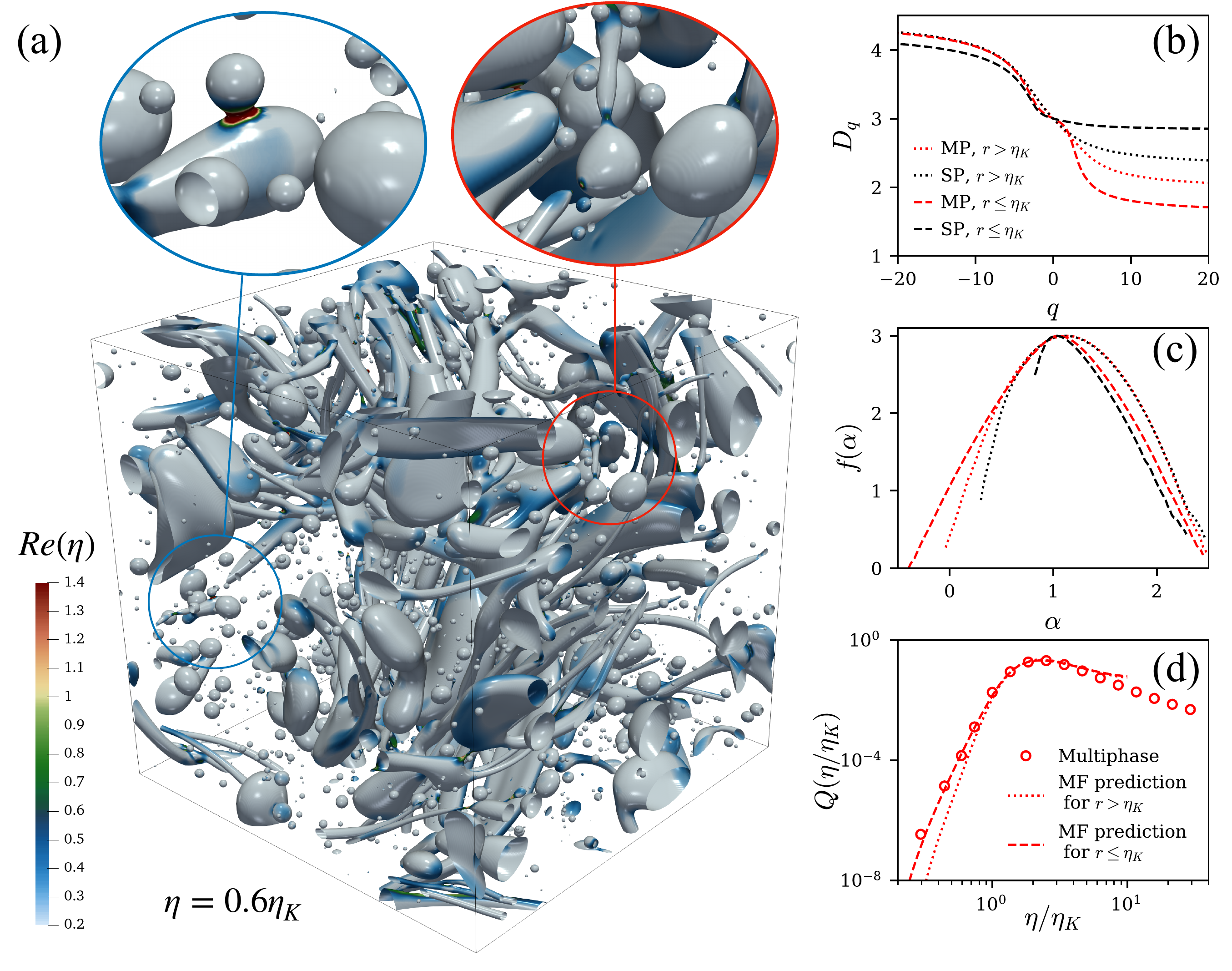}

	\caption{Left (panel a): Iso-surfaces of the color function at $0.5$, showing the liquid--liquid interface and colored by the local Reynolds number $\mathrm{Re}(\eta)$ for $\eta/\eta_K \approx 0.6$. Enlargements highlight representative coalescence (blue) and breakup (red) events. 
    Right: multifractal analysis for the single-phase (SP) and multiphase (MP) simulations.
    (b) Generalized R\'enyi dimensions $D_q$. Dotted lines correspond to the AK range, $2.4<r/\eta_K<20$, and dashed lines to the BK range, $0.3<r/\eta_K<2.4$. (c) Corresponding singularity spectra $f(\alpha)$, using the same convention. (d)~Probability density function $Q(\eta)$ for the multiphase case. Points are the raw DNS data (same as \Cref{fig:qeta}). Lines correspond to the multifractal prediction (\Cref{eq:fitBife}) using  multifractal spectra shown in (c) for the MP case. We use $\mathrm{Re} = 150$, $d=1$ and $A=4.73$ for the sub-Kolmogorov case and $A=4.71$ for the super Kolmogorov case. 
    }
	\label{fig:local_re_multifrac}
\end{figure*}

To address the first point, we examine the instantaneous scale-dependent local Reynolds number $\mathrm{Re}(\textbf{x}, \eta)$ defined in \Cref{eq:localRe}. 
As shown in \Cref{fig:local_re_multifrac}(a), the events contributing to the sub-Kolmogorov tail of $Q(\eta)$ are strongly concentrated around topology-changing interfaces
{notably breakup and coalescence}. The enlarged views further highlight  that these regions are associated with large interface curvature and 
extreme small-scale velocity fluctuations. Remarkably, this picture is found to be robust at different sub-Kolmogorov scales, as shown in the supplemental material. 
By contrast, no comparable concentration of events can be observed in the single-phase case, since this mechanism is specific to multiphase turbulence.
This local analysis demonstrates that the most intermittent fluctuations are spatially organized around breakup and coalescence events. The key question is then whether such localized structures remain a purely interfacial feature, or whether they reshape the statistical geometry of the dissipation field more broadly. Since intermittency is intrinsically tied to the spatial organization of dissipation, these observations suggest that the multifractal properties of multiphase turbulence should differ from those of single-phase flows.

To test this idea, we perform a multifractal analysis of the dissipation field, with the aim of quantifying how interfacial dynamics modify the geometrical and statistical structure of intermittency in multiphase turbulence.
We interpret $\varepsilon(\mathbf{x},t)$ as a singular measure supported in a three-dimensional domain~\cite{Meneveau1987,Mukherjee2024}, and define the dissipated energy coarse-grained over boxes $\mathcal{B}_r$ of size $r$,
$
P_r(\mathbf{x})=\int_{\mathcal{B}_r(\mathbf{x})}\varepsilon(\mathbf{x}')\,d\mathbf{x}'.
$
Since the condition $Re(\eta)\sim 1$ identifies the scales at which viscous regularization becomes locally active, the broadening of $Q(\eta)$ in the multiphase case suggests a modified spatial organization of the dissipation measure. To quantify this organization, we introduce the normalized box measure
$
\mu_i(r)={P_r(\mathbf{x}_i)}/{\sum_{j=1}^{N_r}P_r(\mathbf{x}_j)},
$
and the partition function
$
Z_q(r)=\sum_{i=1}^{N_r}\mu_i(r)^q \sim r^{\tau(q)}$, where $N_r$ denotes the number of boxes of size $r$ and $\textbf{x}_i$ the location of the i$^{th}$ box.
The generalized R\'enyi dimensions are then defined by
$
D_q={\tau(q)}/({q-1}),
$
and the corresponding singularity spectrum is obtained through the Legendre transform
\begin{equation}
\alpha(q)=\frac{d\tau(q)}{dq}, \qquad
f(\alpha)=q\alpha-\tau(q).
\end{equation}
A crucial step is the identification of the scale intervals over which $Z_q(r)$ exhibits approximate power-law scaling. In single-phase turbulence, multifractal scaling is usually sought between the Kolmogorov scale $\eta_K$ and the inertial range~\cite{Meneveau1987,Sreenivasan1991}. Here, however, \Cref{fig:qeta} shows that in the multiphase flow the support of dissipative events extends well below $\eta_K$. We therefore determine the multifractal statistics over two distinct intervals,  $0.3<r/\eta_K<2.4$ and  $2.4<r/\eta_K<20$, hereafter denoted below Kolmogorov (BK) and above Kolmogorov (AK), respectively.
The AK interval corresponds to the range commonly adopted in three-dimensional single-phase analyses~\cite{Mukherjee2024}, while BK probes the near- and sub-Kolmogorov regime where interfacial effects are the strongest.

The resulting $D_q$ and $f(\alpha)$ are shown in \Cref{fig:local_re_multifrac}(b,c). We first analyse the single-phase case for comparison and to assess the whole procedure. In single phase turbulence, the AK analysis recovers the expected singularity spectrum (see also fig.~1 in SM), whereas the BK spectrum is markedly narrower and depleted in low-$\alpha$ events. Thus, the multifractal signature associated with inertial-range intermittency progressively weakens once the analysis is restricted to scales of order $\eta_K$ and smaller. The same trend is visible in $D_q$: at large positive $q$, which emphasizes the most intermittent regions, the AK dimensions are substantially smaller than their BK counterparts.

The multiphase case exhibits a qualitatively different behavior. The spectrum is more singular in any case with respect to the single-phase. Strong left tails persist in both AK and BK, and are particularly pronounced in BK, where very small values of $\alpha$ remain supported on sets with finite fractal dimension. This shows that interfacial dynamics do not merely transmit inertial-range intermittency to smaller scales, but sustain a strongly multifractal dissipative regime in the vicinity of and below $\eta_K$. By contrast, over moderate and large $\alpha$ in AK the multiphase and single-phase spectra are indistinguishable. Consistently, the $q<0$ branches of $D_q$ nearly collapse in AK, indicating that weakly dissipative regions are only mildly affected by the interface, whereas the strongest modifications concern the intense events emphasized by large positive $q$.

To close the loop, following the derivation from \textcite{Biferale2008} originally developed for single phase turbulence, and reproduced in SM, one can infer from the multifractal spectrum, $f(\alpha)$, the distribution of cut-offs lengths
\begin{align}
    Q({\eta}) =&  \frac{1}{3}\int \mathbf{d} \alpha\,A^{x(\alpha)}\mathrm{Re}^{y(\alpha)}{\eta}^{z(\alpha)} \notag \\
&\quad\times \exp\!\left[-A^{2(1-\alpha/3)}\mathrm{Re}^{(\alpha-1)/2}{\eta}^{-2(1+\alpha/3)}\right],
\label{eq:fitBife}
\end{align}
where $A$ is an adjustable parameter of order 1, $\mathrm{Re}$ is the large scale Reynolds number, $x(\alpha)=d(1-\alpha/3)+3- f(\alpha)$, $y(\alpha) =[d(\alpha - 1 )-3(3 - f(\alpha))]/4$, $z(\alpha) = -d(1+\alpha/3) - 1 + (3 - f(\alpha))$ and $d$ is the number of components of the velocity field used to compute the spectrum.
In  \Cref{fig:local_re_multifrac}(e), we show the resulting curves using either BK or AK spectra (see Figure 2 in SM for the single phase counterpart). The agreement is excellent when the BK spectrum is used, while some dissipative events are missed in the sub-Kolmogorov range when using the AK spectrum. Interfaces modify the small scale organization of the velocity field and henceforth of dissipation. Therefore the multifractal approach has predictable powers provided scales are adequately chosen. 

\begin{figure}
	\centering
	\includegraphics[width=0.98\linewidth]{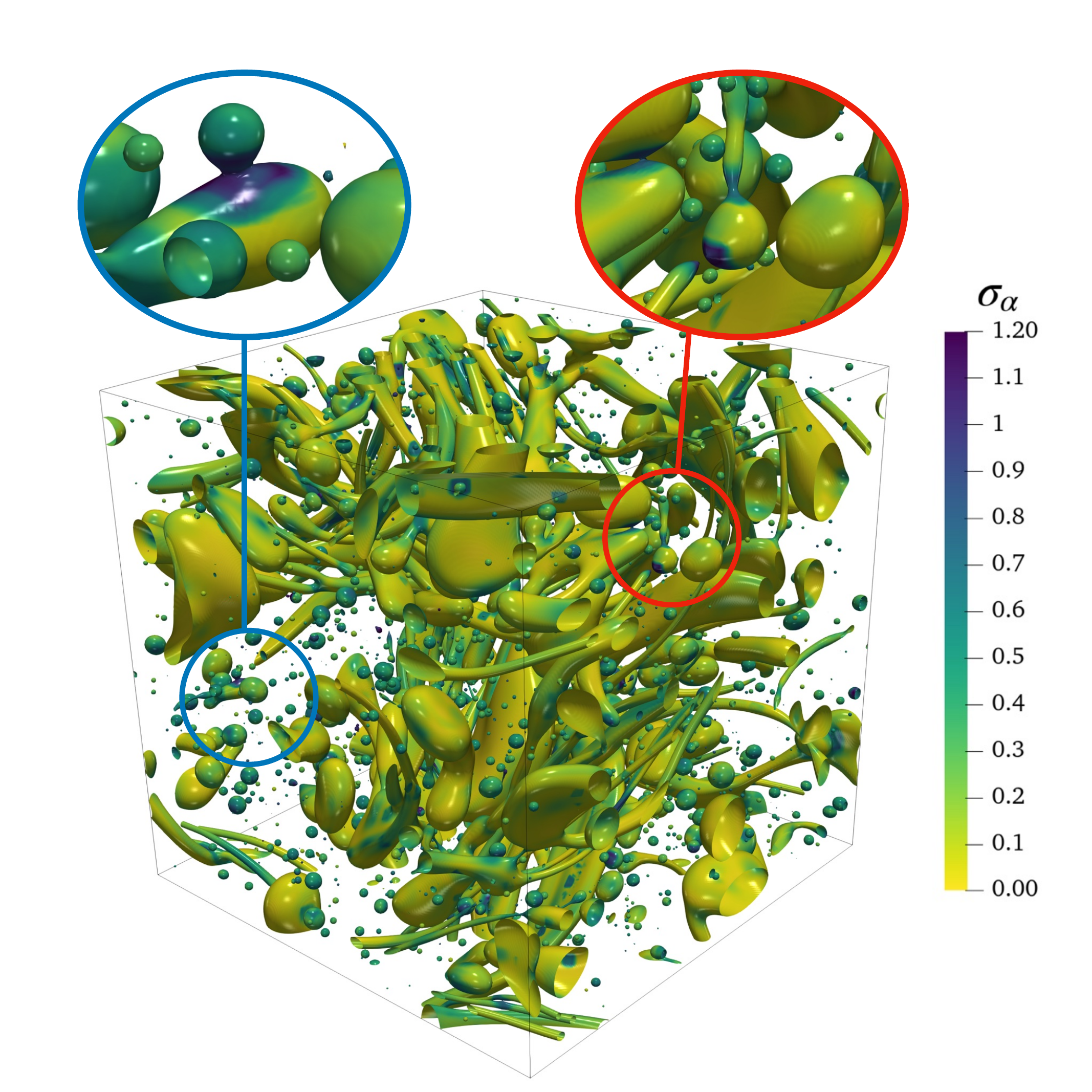}
	\caption{Same visualization as in \Cref{fig:local_re_multifrac}, but colored by the local spread of H\"older exponents, $\sigma_\alpha$. The zoomed regions coincide with those shown in \Cref{fig:local_re_multifrac}.
    }
	\label{fig:local_alphstd}
\end{figure}

A striking feature of the multiphase BK spectrum is the finite value $f(\alpha\approx 0)\approx 1$, suggesting that the most singular dissipative structures are consistent with support on approximately one-dimensional sets. This interpretation is consistent with \Cref{fig:local_re_multifrac}, where the strongest sub-Kolmogorov events cluster around thinning ligaments before breakup and around reconnecting liquid threads during coalescence. Although the underlying topology-changing dynamics cannot be resolved as true singularities, it is nevertheless informative to identify numerically the regions that contribute most strongly to the broadening of the BK spectrum.
To localize these structures, we follow the recent box-based visualization strategy of \citet{Mukherjee2024}, building on earlier attempts to map local H\"older exponents in turbulent flows~\cite{Nguyen2019,Dubrulle2022}. We perform a local multifractal analysis on boxes of size $r/\eta_K=2.4$ centered at each grid point, and define the local spread of H\"older exponents as
$
\sigma^2_\alpha=\left\langle (\alpha-\bar{\alpha})^2 \right\rangle,
$
where $\bar{\alpha}$ denotes the exponent computed averaging over the entire box. Because this procedure probes the BK interval locally, $\sigma_\alpha$ highlights the regions that contribute most to the small-scale broadening of the multiphase spectrum. The resulting field is shown in \Cref{fig:local_alphstd}.\\


As anticipated, the largest values of $\sigma_\alpha$ are concentrated around topology-changing interfacial events, namely breakup and coalescence. Elevated $\sigma_\alpha$ is also observed along thin ligaments and around small droplets, where large curvature localizes the surface-tension force and enhances small-scale velocity gradients.

\paragraph*{Conclusions ---}
We have extended the multifractal approach to multiphase turbulence. We have shown that the small-scale dynamics of multiphase turbulence are characterized by a strong broadening of the local dissipative cutoff, with dissipative events extending deep into the sub-Kolmogorov range. By combining the statistics of $Q(\eta)$ with a multifractal analysis of the dissipation field, we demonstrated that these extreme events are spatially concentrated around topology-changing interfacial regions, namely breakup and coalescence, and give rise to a markedly broader singularity spectrum at small scales. While the multifractal properties above $\eta_K$ remain close to those of single-phase turbulence, the near- and sub-Kolmogorov range exhibits a distinct dissipative geometry, with the strongest events supported on sparse structures associated with interfacial dynamics. We have also provided evidence of the predictive capability of the multifractal approach in multiphase turbulence, assessing the theoretical prediction of the distribution of the dissipative scale, once given the singularity spectrum. Altogether, these results establish that breakup and coalescence do not merely perturb turbulence locally, but imprint a distinct multifractal geometry on the dissipation field.

\begin{acknowledgments}
The authors acknowledge the support of the French Agence Nationale de la Recherche (ANR), under grant SPEED (ANR-20-CE23-0025-01) and SCALP (ANR-24-CE23-1320). AR acknowledges the support from the Swiss National Foundation through the SNSF Swiss Postdoctoral Fellowship, project number 233920. {This project was provided with computing HPC and storage resources by GENCI at TGCC thanks to the grant A0172B10759 on the supercomputer Irene's ROME partition }. MCE acknowledges the support of Fondo di Ateneo per la Ricerca Dipartimentale 2024 (FARD-2024) at University of Modena and Reggio Emilia. MCE also  acknowledge the CINECA award under the ISCRA initiative IsB28\_FGWW, for the availability of high performance computing resources and support.
\end{acknowledgments}

\bibliography{biblio}

\end{document}


\title{Supplemental Material: Sub-Kolmogorov Intermittency and Multifractal Dissipation in Multiphase Turbulence}

\author{Marco Crialesi Esposito}
 \affiliation{DIEF, University of Modena and Reggio Emilia, via Vivarelli 10, Modena 41125, Italy}
\author{Aliénor Rivi\`ere}%

\affiliation{Sorbonne Universit\'e, CNRS, UMR 7190, Institut Jean Le Rond d'Alembert, F-75005 Paris, France}
\affiliation{Laboratory of Fluid Mechanics and Instabilities, Ecole Polytechnique F\'ed\'erale de Lausanne, CH-1015 Lausanne, Switzerland}

\author{Sergio Chibbaro}
\affiliation{Universit\'e Paris-Saclay, CNRS, UMR 9015 LISN, AO team, Orsay Cedex 91405, France}
\affiliation{Inria Saclay - Tau team, B\^at 660 Universit\'e Paris-Saclay, Orsay Cedex 91405, France}

\maketitle
\section{Code details of the large-scale forcing}

We perform Direct Numerical Simulations (DNS) of the incompressible Navier--Stokes equations with surface tension,
\begin{equation}
    \rho(\partial_t u_i + u_j \partial_j u_i)
    = -\partial_i p + \mu \partial_{jj} u_i + \sigma \delta_\sigma \kappa n_i + f_i,
    \label{eq:stNS}
\end{equation}
together with the incompressibility condition $\partial_i u_i = 0$.
Here, $u_i(\mathbf{x},t)$ denotes the velocity field, $p(\mathbf{x},t)$ the pressure, $\rho$ the density, and $\mu$ the dynamic viscosity. In the present work, the density is set to unity and taken identical in both phases, so that $\mu$ coincides numerically with the kinematic viscosity. The surface-tension contribution is given by $\sigma \delta_\sigma \kappa n_i$, where $\sigma$ is the surface-tension coefficient, $\kappa(\mathbf{x},t)$ the local interface curvature, $n_i(\mathbf{x},t)$ the unit normal to the interface, and $\delta_\sigma$ a Dirac distribution localizing the force at the interface. Further details on the numerical implementation are given in \cite{Crialesi-Esposito2023-Flutas}.

The forcing term $f_i$ is used to sustain statistically stationary turbulence. In turbulent emulsions at high volume fraction, forcing restricted to the largest scales, for example at wavenumber $k=1$, is known to promote long-lived large-scale liquid structures containing a substantial fraction of the dispersed phase volume~\cite{komrakova2015numerical,Crialesi-Esposito2022}. This effect is commonly associated with the persistence of spatial symmetries in deterministic forcings such as the Arnold--Beltrami--Childress (ABC) forcing, which is built from a superposition of harmonic modes.

To maximize scale separation while suppressing such persistent large-scale liquid structures, we use a time-dependent variant of the ABC forcing,
\begin{equation}
\label{eq:abc_forcing_components}
\begin{aligned}
f_x &= A \sin (z+\omega t) + C \cos (y+\omega t), \\
f_y &= B \sin (x+\omega t) + A \cos (z+\omega t), \\
f_z &= C \sin (y+\omega t) + B \cos (x+\omega t),
\end{aligned}
\end{equation}
where the spatial coordinates are dimensionless and vary in $[0,2\pi]$, and where we set $A=B=C=1$.

To decorrelate the large-scale forcing, we choose the angular frequency as $\omega = 2\pi/\tau$, with $\tau$ smaller than the Kolmogorov time scale $\tau_K$. In the present study, we set $\tau = 0.8\,\tau_K$. This choice ensures that the forcing decorrelates rapidly enough to avoid interference with the near- and sub-Kolmogorov dynamics that are the focus of this work. As shown in the main text (Figs.~2 and 3), no persistent large-scale liquid structures are observed in the simulations.\\

\section{Multifractal formalism in single-phase turbulence}

We recall in this section the application of the multifractal formalism to predict the whole shape of the probability density function (pdf) of the dissipative scale, $\eta$, for the single-phase case. This was put forward by~\textcite{Biferale2008}, based upon the previous work by \textcite{paladin1987anomalous,paladin1987degrees}.
Then, we show the curves obtained by us through numerical simulations and by extracting the empirical multifractal spectrum.

Turbulence is characterized by non-Gaussian velocity fluctuations on a wide range of scales and frequencies. In the inertial range, i.e., for scales much smaller than the stirring length and much larger than the typical dissipative scale, $\eta \ll r \ll L$,  inertial range statistics in isotropic and homogeneous turbulence is characterized by an anomalous power-law scaling, typically measured in terms of structure functions (SFs):
\begin{equation}
S^{(p)}(r)=\bigl\langle |u(x+r)-u(x)|^p \bigr\rangle
=\bigl\langle |\delta_r u|^p \bigr\rangle
\sim \left(\frac{r}{L}\right)^{\zeta(p)},
\end{equation}
where averages $\langle\cdot\rangle$ encode ensemble averages. Note that, for simplicity, we do not consider possible different scalings between longitudinal and transverse increments relevant here~\cite{buaria2023saturation}.

The signature of an anomalous power law is in the deviations of the scaling exponents from the dimensional, Kolmogorov-like, prediction $\zeta_{K}(p)=p/3$. A powerful and simple phenomenological way to understand intermittency was proposed by Parisi and coworkers, the so-called multifractal formalism (MF)~\cite{benzi1984multifractal}. According to the MF model, Eulerian velocity increments at inertial scales can be characterized by a local H\"older exponent $h$, i.e., $\delta_r v \sim (r/L)^h$, whose probability is $P_h(r)\sim (r/L)^{3-D(h)}$, the function $D(h)$ being the fractal dimension of the set where $h$ is observed. In terms of this description, anomalous scaling is easily recovered by a saddle-point estimate in the limit $r/L\to 0$ (but keeping $r \gg\eta$)~\cite{frisch1996turbulence}:
\begin{equation}
S^{(p)}(r) \propto \int \mathrm{d} h\left(\frac{r}{L}\right)^{ph+3-D(h)}
\sim r^{\zeta(p)},
\tag{2}
\end{equation}
where $\zeta(p)=\min_h[ph+3-D(h)]$. 
Besides SF scaling, the MF formalism proved to be able to predict multiscale correlation functions and probability density functions (pdfs) of velocity differences and velocity gradients for both Eulerian and Lagrangian statistics~\cite{benzi1991multifractality,chevillard2003lagrangian,Boffetta2008}. 
Several phenomenological proposals have been made for the shape of the $D(h)$ spectrum which fits the experimental and numerical data better (random beta model, $p$ model, log-normal, log-Poisson, etc.)~\cite{frisch1996turbulence,dubrulle2019beyond}.
From first principle, because we lack a derivation from the Navier-Stokes equations, they are all on the same footing, except for log-normal models that are known to be affected by some ~\cite{frisch1996turbulence}.

To predict the probability density function (pdf) of the dissipative scale, one uses the well known assumption that it is itself a fluctuating quantity, defined by the requirement that the local Reynolds number is of order 1~\cite{paladin1987degrees,paladin1987anomalous}:
\begin{equation}
\frac{|\delta_{\eta} u|\,\eta}{\nu}\sim 1,
\end{equation}
where $\delta_{\eta}u=\delta_r u$, calculated at scale $
\eta$. 
For the sake of completeness, we provide here the developments as put forward by Biferale~\cite{Biferale2008}.
The argument is as follows. Suppose you have a large-scale Gaussian velocity field with $d$ components, $u_0$, with order 1 variance, $u_{\mathrm{rms}}=1$. Its amplitude must have the following pdf:
\begin{equation}
P(u_0) d u_0 = u_0^{d-1}\exp(-u_0^2/2)d u_0,
\end{equation}
where we have neglected unessential order unity prefactors. Now, let us define the fluctuating dissipative scale by the Paladin-Vulpiani argument, requiring $O(1)$ local Reynolds number,
\begin{equation}
\frac{\eta(h,u_0)|\delta_{\eta}v|}{\nu}\sim 1;
\end{equation}
using the multifractal scaling,
\begin{equation}
|\delta_{\eta}u|=u_0\left(\frac{\eta}{L}\right)^h
\end{equation}
 we get for the expression which connects $\eta$ to $u_0$
\begin{equation}
u_0=Re^{-1}\left(\frac{\eta}{L}\right)^{-(1+h)},
\end{equation}
where the Reynolds number is defined as $Re=Lu_0/\nu$, the typical amplitude of $u_0$ being $v_{\mathrm{rms}}=1$.

Let us consider the dimensionless dissipative scale,
\begin{equation*}
{\eta}=\frac{\eta}{\eta_K},
\end{equation*}
where the Kolmogorov scale is defined as $\eta_{K}=L Re^{-3/4}$. In terms of a dimensionless variable, we have 
\begin{equation}
u_0=Re^{(3h-1)/4}{\eta}^{-(1+h)}.
\end{equation}
Now, conservation of probability implies (for each given $h$ exponent)
\begin{equation}
\phi({\eta}|h)=P(u_0)\left(\frac{d u_0}{d{\eta}}\right).
\end{equation}
Consider also the fluctuations of $h$ and that the probability to measure an $h$ exponent at scale ${\eta}$ is given by the multifractal rule $P_h(\eta)=(\eta/L)^{3-D(h)}$. Then, we obtain
\begin{equation}
    \phi({\eta}) \propto \int d h\,P_h(\eta)\phi({\eta}|h) = \int d h\,Re^{y(h)}{\eta}^{z(h)}
\exp\!\left[-0.5\,Re^{(3h-1)/2}{\eta}^{-2(1+h)}\right],
\label{eq:prob}
\end{equation}
where we have defined $y(h)=[(3h-1)d-3(3-D(h))]/4$, and $z(h)=-d(1+h)-1+(3-D(h))$. Notice that for K41 scaling [i.e., the whole $h$ support limited to $h=1/3$, with $D(1/3)=3$], we recover that the pdf of a dimensional dissipative scale does not depend on Reynolds (K41 does not have any dependency on Reynolds for dimensionless variables).
This is the multifractal prediction.

The requirement to have an order unity local Reynolds can be further relaxed by taking into account that velocity fluctuations are not an exact pure power law from the large scale down to the local dissipative scale. In other words, one can introduce a smooth transition between the inertial range behavior and the dissipative behavior (for each $h$ exponent). This can be done using a generalization of the Batchelor parametrization~\cite{meneveau1996transition}
\begin{equation}
\delta_r u = u_0\frac{r/L}{\bigl((r/L)^2+c(\eta/L)^2\bigr)^{(1-h)/2}},
\label{eq:batch}
\end{equation}
where $c$ is an $O(1)$ free parameter defining the matching between the two regimes. Relation (\ref{eq:batch}) together with the corresponding generalization for the probability to observe a given $h$ exponent: $P_h(r)\propto ((r/L)^{3-D(h)}+c(\eta/L)^{3-D(h)})$, leads to a consistent, simultaneous description of both inertial and dissipative range physics. Notice that the above receipt is also consistent with the requirement $\lim_{Re\to\infty}\langle(\partial v)^2\rangle\sim Re$, as requested by the existence of the dissipative anomaly~\cite{frisch1996turbulence}. By taking into account this extra degree of freedom, one gets a slightly modified version of (\ref{eq:prob}) when using definition (\ref{eq:batch}):
\begin{equation}
    \phi({\eta}) = \int d h\,P_h(\eta)\phi({\eta}|h) = \int d h\,A^{x(h)}Re^{y(h)}{\eta}^{z(h)} 
\quad\times \exp\!\left[-A^{2(1-h)}Re^{(3h-1)/2}{\eta}^{-2(1+h)}\right],
\end{equation}
where $A=(1+c)^{1/2}$ and $x(h)=d(1-h)+3-D(h)$. 
The multifractal prediction is derived from velocity increments, and then we have used the Holder exponent $h$ for the scaling. In our work we have considered the dissipation field and the related scaling exponent is denoted by $\alpha$. Using the standard bridge relations between velocity increments and dissipation \cite{frisch1996turbulence}, we obtain that $\alpha=3h$ is used in the main article.

The shape of the dissipative scale $\phi(\eta)$ has been also proposed~\cite{Schumacher2007a} using a formulation based on the Mellin transform of the structure functions and the assumption that the velocity statistics is Gaussian at large scale.  The results are similar to those obtained in~\cite{Biferale2008}, where the $D(h)$ was prescribed to be of the She-Leveque form~\cite{She1994}.

\section{Results in the single-phase turbulent flow}
\begin{figure}[h]
    \centering
    \includegraphics[scale=0.8]
    {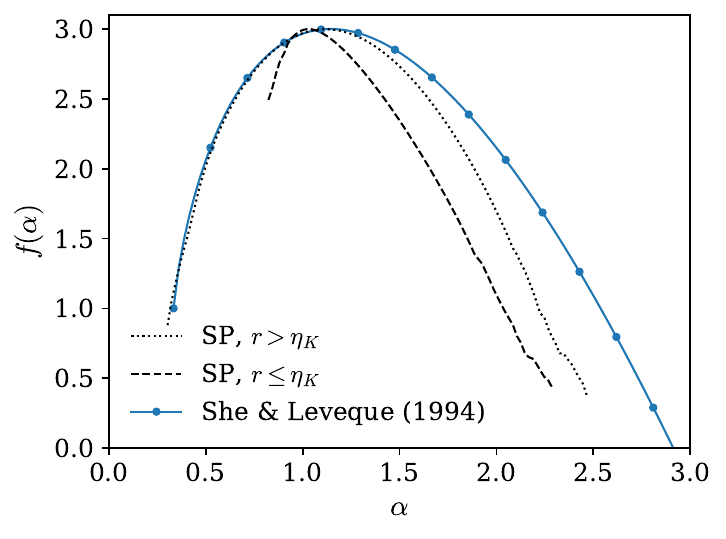}
    \caption{Singularity spectra $f(\alpha)$ of single phase turbulent flows for the two ranges AK and BK, compared to the model proposed by \textcite{She1994}.}
    \label{fig:falpha-SheLeveque}
\end{figure}
In figure \ref{fig:falpha-SheLeveque}, we show our results concerning the multifractal spectrum $f(\alpha)$ for the case of single-phase turbulence together with She-Leveque theoretical model predictions for comparison.

In particular, we repeat the procedure used in the main work, using the multifractal spectrum $f(\alpha)$, as extracted from simulations in two different ranges, above  and below the Kolmogorov scale.
We can appreciate that the spectrum computed above Kolmogorov is in excellent agreement with the She-Leveque one, which is known to fit quite well the scaling exponent of single-phase turbulence, up to $\alpha\approx 1.5$. Larger values are mostly controlled by large-scales which are not resoleved in our simulations. That explain the difference. These values are related to smooth parts of the flow, and are not of interest in the present work.
When the spectrum is computed using the scales below Kolmogorov, the shape is different with a much reduced variance. That shows, as expected, that in single-phase turbulence fluctuations tend to decrease in the far-dissipation range.

\begin{figure}[h]
    \centering
    \includegraphics[scale=0.8]{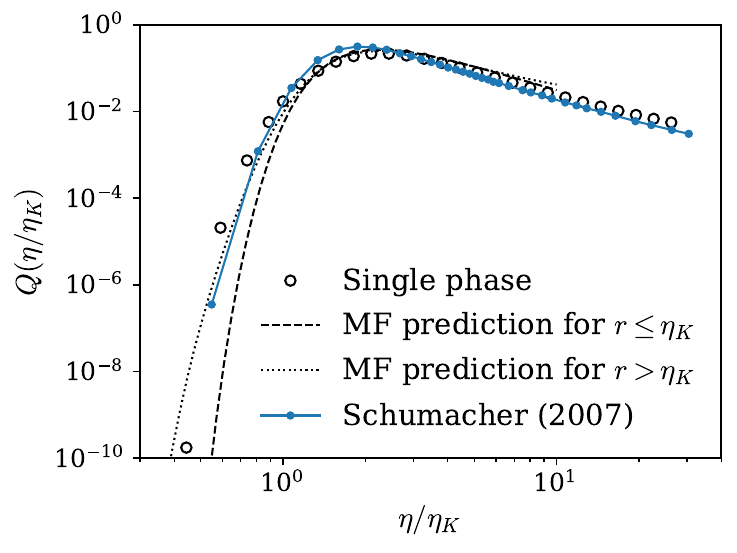}
    \caption{Distribution of dissipative scales $\eta/\eta_K$ obtained single phase simulations (black dots) compared to the dataset from \textcite{Schumacher2007a}. Lines correspond to the fit using eq. 3 of the main text, with the multifractal spectrum for the 2 ranges AK, BK, shown ~\ref{fig:falpha-SheLeveque}. We use $\mathrm{Re} = 150$, $d=1$ and $A=4.58$ for the sub-Kolmogorov case and $A=4.66$ for the super Kolmogorov case. }
    \label{fig:Qeta-sp}
\end{figure}

In figure \ref{fig:Qeta-sp}, we show the distribution of dissipative scales for the single-phase turbulence case, to complement the figure 1 of the main manuscript.
The distribution computed from our simulations are directly compared against previous DNS by Schumacher~\cite{Schumacher2007a}. 
The two DNS are in very good agreement, with our simulations having a slightly higher resolution.
Then, as in the main manuscript for the multiphase case, we compare against the multifractal prediction. We use both the spectrum extrcated above and below the Kolmogorov scale.
The agreement is excellent for the left tail between DNS and the multifractal prediction using the scales above Kolmogorov. Instead, the scales below the Kolmogorov scale produce much less fluctuations, consistently with the previous results.

Overall, the results show that our simulations are fully consistent in single-phase turbulence, and they corroborate the consistency of our method to extract the multifractal spectra.

\section{Topological events in the far dissipative range}
\begin{figure}[h]
    \centering
    \includegraphics[width=0.7\linewidth]{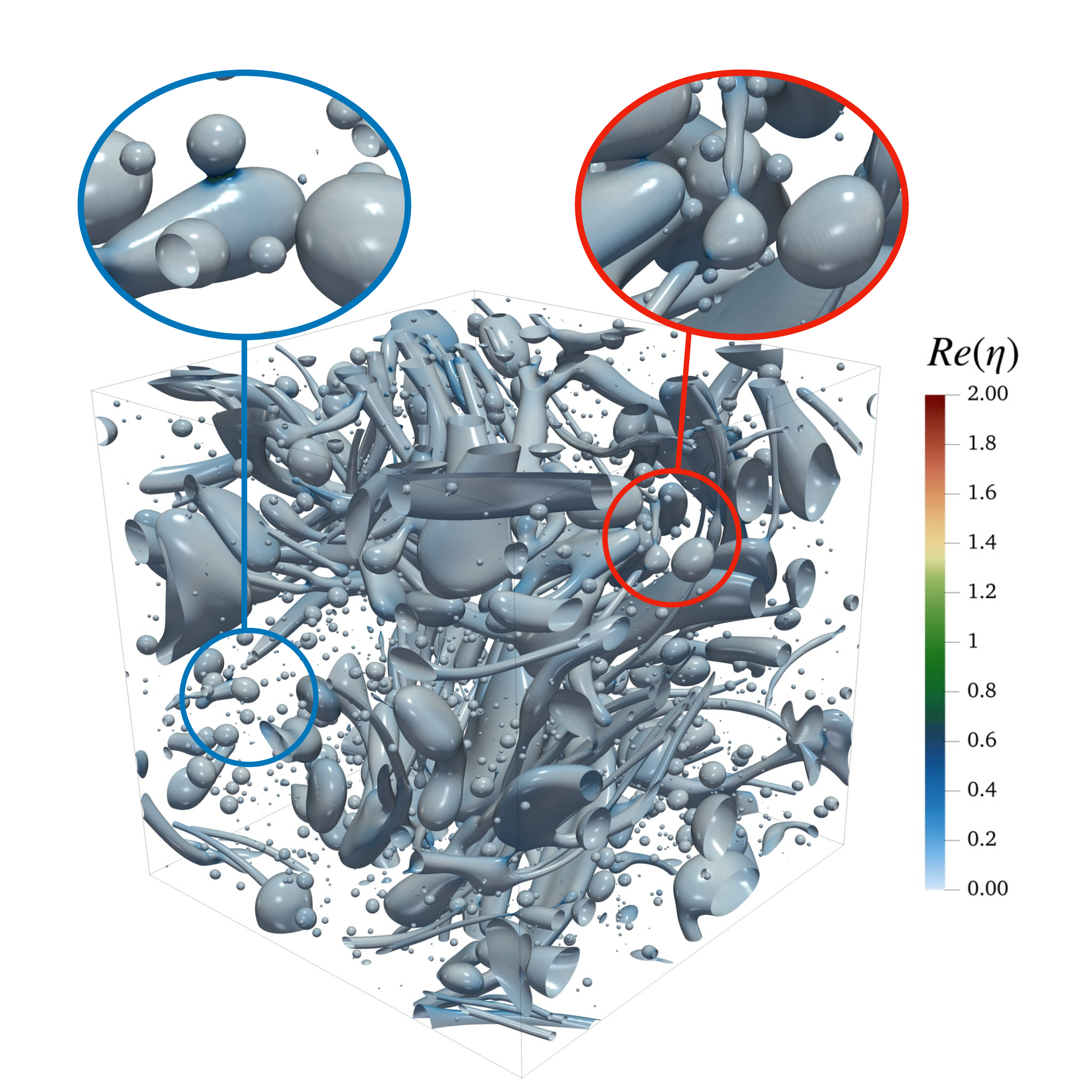}
    \caption{Iso-surfaces of the color function at $0.5$, showing the liquid--liquid interface and colored by the local Reynolds number $\mathrm{Re}(\eta)$ for $\eta/\eta_K \approx 0.3$ }
    \label{fig:rende_supp}
\end{figure}

In figure \ref{fig:rende_supp}, we complement the results shown in Fig. 2 of the main manuscript with another render obtained at a scale even smaller, namely for $\eta=0.3 \eta_K$. 
As expected, the figure points to less extreme events than at larger scales, and yet it maintains the same qualitative behavior indicating that large dissipative fluctuations are concentrated near topological interface changes.
That emphasize the universality of our results in the intermediate-dissipative range studied.

\bibliography{biblio}